# Spontaneous decoherence as a result of randomness in the flow of time[1]

L.I. Rozonoer[*]

Institute of Control Sciences of the Russian Academy of Sciences

### Abstract

The hypothesis of the random flow of time is considered. To do this, the concepts of microscopic random time and macroscopic mean time, as well as random modular time are introduced. The possibilities of experimental verification of the hypothesis of the random flow of time are discussed. The possibility of interpreting the time-like fifth coordinate (introduced by V.A. Fock) as absolute time is considered and the question of the reality of absolute time is discussed. The absolute time, which manifests itself as the proper time of the particles, is assumed to be random.

**1. Introduction.** It is now widely accepted that quantum behavior transforms into classical behavior as a result of decoherence of quantum states (see for example [1]). Decoherence is the basis of the quantum theory of measurement (see, e.g., survey [2]). The overwhelming majority of physicists consider decoherence to be the result of a connection between a quantum system and its environment (*environment-induced decoherence*). The existence of the environment-induced decoherence is supported theoretically (see for instance [1]) and by experiments with open quantum systems (for example, in [3] decoherence was induced by coupling a system to an engineered reservoir). However, these experiments aimed at investigating environment-induced decoherence, and therefore they cannot eliminate a possibility that decoherence can also exist in isolated quantum systems (*spontaneous decoherence*).

A hypothesis suggesting that spontaneous decoherence can be a result of a "spontaneous localization" (random collapses of the wave function) was proposed by Ghirardi, Rimini, Weber and others (see, e.g., [4]). Later, various models of the collapse were worked out (see exhaustive surveys in [5], [6], and the discussion of a possible experiment in [7]).

Spontaneous decoherence of quantum states and the violation of the unitarity of a quantum system's evolution are the subject of numerous recent works. The majority of these works attribute these effects to quantum gravity (see survey [8]). Paper [9] considers and rejects some common arguments against spontaneous decoherence.

Works [10], [11] considered spontaneous decoherence in discrete models of quantum gravity and discussed possible experiments.

Papers [12-14] studied decoherence in an isolated system caused by the fluctuations of classical parameters, which determine the time evolution of the system (for example, the fluctuations of the duration of the energy levels' evolution). The authors developed a general formalism describing this type of decoherence (which they called *non-dissipative decoherence*), and applied this formalism to trapped ions and quantum computers. According to the authors, in these cases non-dissipative decoherence dominated over environment-induced decoherence.

Another approach to decoherence was developed in [15], where decoherence emerged if the evolution of a quantum system was measured with a clock impacted by random fluctuations.

Paper [16] proposed a modification of quantum mechanics, where the evolution was described not by a single continuous unitary transformation, but by a stochastic sequence of identical unitary transformations.

Finally, a comparison of different approaches (including [16]), which attribute decoherence to time randomness, is made in [17].

Below I introduce the hypothesis that spontaneous decoherence is caused by randomness in the flow of time. It must be stressed that this mechanism differs conceptually from those considered in [12-14] and [15]. Non-dissipative decoherence (in the sense of the works [12-14]) does not assume that the unitarity of evolution is violated: each single process of evolution is unitary, and decoherence appears as a result of averaging of these processes over given distribution of the parameters. Neither is the result of unitarity violation the type of decoherence considered in [15]; it rather appears as a result of uncertainties of the time measurements. In contrast, decoherence induced by randomness in the flow of time violates unitarity, similarly to the random collapse models. However, as it will be shown below, the equations describing evolution with randomness in the flow of time do not violate any conservation laws, while in the collapse models the energy conserves only approximately.

We will see later that the model of decoherence developed in [16] is a special case of decoherence caused by randomness in the flow of time.

My motivation for studying spontaneous decoherence, firstly, is based on the intention to understand the transformation of a pure state into a mixed state in a quantum measurement. Secondly, it is my hope to overcome, -- by introducing a fundamental irreversibility into the description of dynamics, -- the logical difficulties (underlined in [18]) which arise when micromechanics is combined with the principles of statistical mechanics. It is shown in [18] that micromechanics (either classical or quantum) is logically incompatible with randomness and cannot contain any source of randomness. The proposed hypothesis identifies randomness in the flow of time as such a source. I hope that spontaneous decoherence, if it exists, can provide a basis for the principles of statistical physics.

*In a wider context, it may appear that randomness in the flow of time could be a trace of the space-time foam imprinted on the classical space-time (a possibility of observing the noise induced by the space-time foam in low-energy experiments was highlighted in [19]). The discussion of the hypothesis of randomness in the flow of time in the context of quantum gravity can be found in the Appendix C.*

**2. Randomness in the flow of time and decoherence.** Let a random time interval $\theta$ be characterized by its mean value (i.e. mean duration) *t*. If the consecutive intervals of time are independent and stationary random variables, then the "*microscopic*" random time $\theta$ is a random stationary process $\theta(t)$ with independent increments [20]*,* considered in the "*macroscopic*" time *t.* For a process of this kind, the standard deviation has a value of the order of $(t\tau)^{\frac{1}{2}}$, where $\tau$ is some characteristic time. Below it is assumed that, in the non-relativistic case, $\tau$ is a universal constant, which determines the scale of randomness of the time flow. At τ=0, the time flow is deterministic. The randomness of the microscopic time leads to randomization of the phase differences and destroys the phase relations between oscillations: the phase difference of two oscillations with a frequency difference $\Delta\omega$ becomes a random quantity with the standard deviation of the order of $\Delta\omega(t\tau)^{\frac{1}{2}}$. After the time interval $t_{dec}$ for which $\Delta\omega(t_{dec}\tau)^{\frac{1}{2}} \geq \pi$ oscillations become incoherent. Therefore a superposition of two stationary states with energy difference $\Delta E$ converts spontaneously to a mixed state after time interval which obeys the inequality (put $\Delta \omega = \frac{\Delta E}{h}$) :

$$\frac{\Delta E}{h}(t_{dec}\tau)^{\frac{1}{2}} \geq \pi .$$

Therefore

$$t_{dec} = \frac{\pi^2 h^2}{\tau(\Delta E)^2} . \qquad (2.1)$$

In the relativistic case, we must consider the Minkowski intervals between the events rather than time intervals. For this reason, the effect of spontaneous decoherence does not occur for propagation of light (the intervals between the events of light observation are equal to zero). For a moving particle,

$$\tau = \frac{\tau_0}{\left(1 - v^2/c^2\right)^{\frac{1}{2}}} . \qquad (2.2)$$

where $v$ is the speed of the particle, $\tau_0$ is the Lorentz invariant.

To modify the conventional quantum dynamics in order to account for random time, we can assume that all physical quantities should be calculated by averaging their conventional quantum expressions over the probability density corresponding to the random process $\theta(t)$, and consider the dynamics in the macroscopic (mean) time $t$ (see next Section). It turns out that for a wide class of random processes $\theta(t)$, the averaged density matrix $R(t)$ is diagonal in the energy representation when $t \to \infty$. Indeed, the non-diagonal elements $R_{kl}, \quad k \neq l$ of the density matrix in the energy representation are proportional to the factors, which exponentially tend to 0 as $t \to \infty$:

$$R_{kl} \approx exp\left[-a\left(\tau/h^2\right)\left(E_k - E_l\right)^2 t\right], \quad a > 0, \qquad (2.3)$$

and the diagonalization of the matrix $R(t)$ occurs on the time scale given by the time interval (2.1) with $\Delta E = E_k - E_l$. This can be viewed as if in this time interval the system measures its own energy [$a$ in (2.3) is a dimensionless factor of the order of 1, whose explicit value depends on the random process $\theta(t)$].

The evolution of a quantum system in the macroscopic time $t$ is irreversible (in spite of the fact that it is described by conventional quantum mechanical equations in microscopic time $\theta$). In particular, the entropy of the system $S = -Tr(R \ln R)$ increases.

**3. Modification of quantum dynamics.** Let a time interval $\theta$ be a random quantity characterized by a mean value $t$ and by a probability density $P(\theta; t)$. If we consider the random process $\theta(t)$ as a process with independent stationary positive increments, we arrive at the generalized Poisson distribution [20]:

$$P(\theta; t) = \exp\left(-\frac{t}{\tau}\right) \sum_n \left(\frac{1}{n!}\right)\left(\frac{t}{\tau}\right)^n \left[\left(\frac{1}{\tau}\right)p\left(\frac{\theta}{\tau}\right)\right]^{(n)*}, \tag{3.1}$$

where $p(\xi)$ is the probability density on the positive half-line, $\xi \geq 0$ with the average $\xi_{av} = 1$, and therefore:

$$\int p(\xi)\,d\xi \;=\; \int \xi\; p(\xi)\,d\xi \;=\; 1, \tag{3.2}$$

the symbol $(n)^*$ denotes the n-fold resultant (for n>1) of a function [with the convention: $p^{(0)*}(\xi) = \delta(\xi), \; p^{(1)*}(\xi) = p(\xi)$], and $\tau$ being a constant. Equation (3.1) can be interpreted in the following way. A clock ticks $n$ times in a mean (macroscopic) time $t$, where $n$ is a random number, which follows the Poisson distribution with the average $n_{av} = \dfrac{t}{\tau}$. The intervals between the ticks, $\Delta$, are independent random quantities having the same probability density $\left(\dfrac{1}{\tau}\right)p\left(\dfrac{\Delta}{\tau}\right)$ with the mean value $\Delta_{av} = \tau$. The characteristic function of the probability density (3.1) is equal to

$$\int P(\theta; t)\exp(i\lambda\theta)\,d\theta = \exp\left[\left(\varphi(\lambda\tau) - 1\right)\left(\frac{t}{\tau}\right)\right], \tag{3.3}$$

where $\varphi(\varsigma)$ is the characteristic function of the probability density $p(\xi)$. For a "conventional" Poisson process,

$$p(\xi) = exp(-\xi), \qquad \varphi(\varsigma) = (1 - i\varsigma)^{-1}. \tag{3.4}$$

To account for time randomness in the equations of quantum dynamics, let's assume that all physical quantities should be calculated by averaging their conventional quantum mechanical expressions over the probability density (3.1), and after the averaging the dynamics should be considered in the "mean time" $t$ (see Appendix A for a discussion of possible techniques for introducing randomness). It means that the values of all physical quantities must be calculated with the help of the averaged density matrix

$$R(t) = \int \rho(\theta) P(\theta;t) d\theta , \qquad (3.5)$$

where $\rho(\theta)$ obeys conventional equation

$$ih \frac{d[\rho(\vartheta)]}{d\theta} = [H, \rho]. \qquad (3.6)$$

Here $H$ is the Hamiltonian, and the brackets [,] stand for the commutator. The equation for the averaged density matrix $R(t)$ can be obtained from the expression

$$\frac{\partial P(\theta;t)}{\partial t} = \left(\frac{1}{\tau}\right)\left[\int P(\theta - \xi\tau;t)p(\xi)d\xi - P(\theta;t)\right], \qquad (3.7)$$

which can be easily verified with the aid of (3.1). Equation (3.7) assumes the convention $P(\theta;t) = 0,$ if $\theta \leq 0$.

We can conclude from (3.6) that

$$\rho(u + \xi\tau) = \exp\left(\frac{-iH\xi\tau}{h}\right)\rho(u)\exp\left(\frac{iH\xi\tau}{h}\right). \qquad (3.8)$$

Let's multiply Eq. (3.7) by $\rho(\theta)$, integrate over $\theta$ and change the variable to $\theta = \eta + \xi\tau$. Then using (3.8) we can write the derivative $\frac{dR(t)}{dt}$ in the following form:

$$\frac{dR(t)}{dt} = \left(\frac{1}{\tau}\right)\left[\int \exp\left(\frac{-iH\xi\tau}{h}\right)R(t)\exp\left(\frac{iH\xi\tau}{h}\right)p(\xi)d\xi - R(t)\right]. \qquad (3.9)$$

Equation (3.9) replaces conventional equation (3.6). It reduces to (3.6) in the limit $\tau \to 0$. Choosing $p(\xi) = \delta(\xi - 1)$, we recover the model considered in [16]. Expanding the right-hand side of (3.9) as a series with respect to $\tau$ and keeping the terms up to $\tau^2$, we get with the help of (3.2):

$$\frac{ihdR(t)}{dt} = [H, R] + a^2 \frac{\tau}{2ih}[H, [H, R]],$$

where $a^2 = \int \xi^2 p(\xi)d\xi$, and $a^2 = 2$, if $p(\xi) = \exp(-\xi)$. For this case we have:

$$\frac{ihdR(t)}{dt} = [H, R] + \frac{\tau}{ih}[H, [H, \rho]].\tag{3.10}$$

Note that equation (3.10) follows directly from (3.5) and (3.6), if $\rho(\theta)$ is averaged in (3.5) with respect to the Gaussian probability density

$$P(\theta; t) = (2\pi\tau)^{-\frac{1}{2}} \exp\left[\frac{-(\theta - t)^2}{2t\tau}\right].\tag{3.11}$$

We can solve equation (3.9) by averaging the solution of equation (3.6), $\rho_{kl} = \exp\left[\frac{-i(E_k - E_l)\theta}{h}\right]\rho_{kl}(0)$, over the probability distribution (3.1) in the energy representation. Using (3.3), we obtain the solution as

$$R_{kl}(t) = R_{kl}(0)\exp\left\{\left[\phi\left(\left(\frac{E_k - E_l}{h}\right)\tau\right) - 1\right]\left(\frac{t}{\tau}\right)\right\},\tag{3.12}$$

where $\varphi(\varsigma)$ is the characteristic function of the probability density $p(\xi)$.

For the Poisson process (3.4) we have:

$$R_{kl}(t) = R_{kl}(0)\exp\left\{\left[1 + \frac{(E_k - E_l)^2\tau^2}{h^2}\right]^{-1}\left[\frac{-i(E_k - E_l)t}{h} - \frac{(E_k - E_l)^2\tau t}{h^2}\right]\right\}.\tag{3.13}$$

Since a negative term, $-\frac{(E_k - E_l)^2\tau t}{h^2}$, stays in the exponent, the element $R_{kl}(t)$ decays if $\frac{(E_k - E_l)(\pi)^{\frac{1}{2}}}{h} >> 1$ (see inequality (2.1) in Section 2). Thus, when time $t$ increases, the density matrix in the energy representation becomes diagonal. In particular, a superposition of two states with fixed energies, $E_k$ and $E_l$, $E_k \neq E_l$, transforms into a mixture of these states within the time interval $\hat{t} > \frac{h^2}{\tau(E_k - E_l)^2}$ (see inequality (2.1)).

The diagonalization of the density matrix fails to happen only if the levels $k$ and $l$ are degenerate, so that $E_k = E_l$. In this case the element $R_{kl}(t)$ does not change at all. The diagonalization of the density matrix in the energy representation can be viewed as if the system measures its own energy. One can say that a "self-measurement" of energy takes place. In this regard the proposed modification of quantum mechanics is similar to the modification discussed in [4], where particles "measure" their own coordinates.

The solution of equation (3.10) has the form (in the energy representation):

$$R_{kl}(t) = R_{kl}(0)\exp\left[\frac{-i(E_k - E_l)t}{h} - \frac{(E_k - E_l)^2 \tau t}{h^2}\right]. \tag{3.14}$$

This expression was obtained by averaging the solution of equation (3.6), i.e. the expression $\rho_{kl}(\theta) = \exp\left[\frac{-i(E_k - E_l)\theta}{h}\right]\rho_{kl}(0)$, over the probability distribution (3.11).

The microscopic time obeying the probability distributions (3.1) and (3.11) will be called the "generalized Poissonian time" and "Gaussian time", respectively. The term "Poissonian time" will be reserved for a Poisson process with the density (3.4)

To avoid misunderstanding, it is important to underline that the conclusion about decoherence of stationary states is correct only for the systems with a discrete energy spectrum. In particular, stating that a free particle in a certain superposition of plane waves will decohere into their mixture would be incorrect. Furthermore, while discussing decoherence of states of observable particles it is necessary to take into account the measuring device, and to consider the whole system "object + device". For that matter, J. von Neumann noted [21] that the state of a system "object + device" after the measurement is a stationary state.

Solutions (3.13) and (3.14) of equations (3.9) and (3.10) have absolutely analogous features. Both equations are irreversible in the sense that the entropy $S = -Tr(R \ln R)$ increases when $t$ increases, unless all energy levels are degenerate, and the density matrix in the energy representation is non-diagonal (see Appendix B). While the non-diagonal elements of the density matrix vanish with the increase of $t$ for both equations, the diagonal elements do not change [one can see it from equations (3.12) and (3.13)]. Note that the above arguments do not hold for the case $p(\xi) = \delta(\xi - 1)$ considered in [16]. If $E_k - E_l = \frac{2\pi h s}{\tau}$, where $s$ is an integer number, then the matrix element $R_{kl}$ does not change with time. For example, for a harmonic oscillator with a frequency $\omega = \frac{2\pi}{\tau}$ the density matrix remains constant at all times ("stroboscopic effect").

The diagonal matrix elements of the density matrix in the energy representation calculated in the basis of the solutions of equations (3.9) and (3.10) do not change with time [this can be seen from the derivation of equations (3.9) and (3.10), or from equations (3.12) and (3.13)]. Hence, the norm of the density matrix conserves. As a result, the energy and the momentum of a free particle conserve as well: conservation of the diagonal elements of a density matrix implies the conservation of all physical quantities with the operators, which are diagonal in this (energy, in our case) representation – i.e. which commute with the Hamiltonian. The conservation of a physical quantity means, of course, that conserves not only its mean value, but also its probability distribution.

An irreversible equation analogous to (3.10) was considered in [22] in the context of a modification of quantum mechanics suitable for discrete time. Similar equations were suggested also in [23] in connection with theory of open quantum systems, and were extensively cited later on.

The transition rates calculated using equations (3.9) and (3.10) do not differ significantly from the results obtained in the framework of conventional quantum mechanics, as long as these probabilities are much smaller than $\frac{1}{\tau}$. For example, the law of exponential decay of a state, $P(t) = \exp\left(\frac{-t}{T}\right)$, for the Poisson case (3.4) is modified as

$$\hat{P}(t) = \exp\left[\frac{-t}{T + \tau}\right]. \tag{3.15}$$

The lifetime $T$ transforms to $T + \tau$, and if $\tau \ll T$, this change is not essential. Averaging with the Gaussian density (3.11) gives the analogous result if $\tau \ll t$ (note that the Gaussian approximation of the probability density for a Poisson random process is meaningful only for $\theta > 0$ and $\tau \ll t$).

In Appendix D we consider some modifications of the generalized Poisson distribution. As an illustration to the ideas presented in this chapter, Appendix E describes the time evolution of a Gaussian wave packet in a random time.

**4. Modification of classical dynamics.** Considering random time in classical dynamics is barely interesting from the physical point of view. However, such consideration reveals important traits of the random time hypothesis, for example, the possibility to extend this hypothesis to relativistic dynamics without assuming space-time randomness (the particle's proper time rather than the space-time itself being random).

The classical equation, which determines the evolution of a probability density $w(\theta)$ in a phase space for the microscopic time $\theta$ is the Liouville equation:

$$i\frac{\partial w(\theta)}{\partial \theta} = Lw(\theta),$$  (4.1)

where $L$ is the Liouville operator. The coordinates *(x,p)* of a phase space point are omitted in the expression $w(x, p, \theta)$ to simplify the notations. Following I. Prigogin (see, e.g., [24]), we introduced the imaginary unit into the Liouville equation, so that the operator $L$ becomes self-adjoint. In complete analogy with the procedure developed in the previous section for the averaged density matrix, we will derive an equation of evolution (in the macroscopic time, t) for $W(t)$, the probability density averaged over the microscopic time $\theta$, defined as

$$W(t) = \int w(\theta)P(\theta;t)d\theta .$$  (4.2)

Here, again, the arguments *(x,p)* are omitted, and the probability density $P(\theta;t)$ of the microscopic time is the generalized Poisson distribution [equation (3.1)]. Now we make use of equation (3.7). Multiplying (3.7) by $w(\theta)$, integrating over $\theta$ (with the change of variable $\theta = \eta + \xi\tau$), and taking into account that, by virtue of (4.1),

$$w(\eta + \xi\tau) = \exp(-i\xi\tau L)w(\eta),$$

we obtain the following equation for the averaged probability density in the macroscopic time:

$$\frac{\partial W(t)}{\partial t} = \left(\frac{1}{\tau}\right)\left[\int \exp(-i\xi\tau L)p(\xi)d\xi - 1\right]W(t).$$  (4.3)

We expand the right-hand side of (4.3) in a power series with respect to $\tau$ and keep the terms up to the order of $\tau^2$. Taking into account (4.2), we get:

$$\frac{\partial W(t)}{\partial t} = \left[-iL - \left(\frac{a^2}{2}\right)\tau L^2\right]W(t),$$

where $a^2 = \int \xi^2 p(\xi)d\xi$.

For the Poisson process, $p(\xi) = \exp(-\xi)$, $a^2 = 2$. Then our approximation gives:

$$\frac{\partial W(t)}{\partial t} = -iLW(t) - \tau L^2 W(t).$$  (4.4)

Note that if the averaging is done using the Gaussian density (3.11), equation (4.4) is exact and follows directly from (4.1) and (4.2).

Equation (4.3) has the same structure as equation (3.9), and equation (4.4) is analogous to equation (3.10). This correspondence is a result of the general principle of relationship between classical and quantum mechanics, because the application of the Liouville operator (with the factor (-*i*), in Prigogin's notations) to a function of canonical variables yields the Poisson bracket of this function and the Hamiltonian:

$$-iLW = \{H,W\} = \sum \left[ \frac{\partial H}{\partial p_k} \frac{\partial W}{\partial x_k} - \frac{\partial W}{\partial p_k} \frac{\partial H}{\partial x_k} \right],$$

and a Poisson bracket of classical mechanics corresponds, in quantum mechanics, to the commutator of the Hamiltonian with the corresponding operator divided by *ih:*

$$\{H,W\} \rightarrow \left( \frac{1}{ih} \right) [H,W].$$

As an example, let us consider a particle in a free motion. If the axis *x* is directed along the constant velocity vector, ***v**=**p**/m*, then in equation (4.4) we can set

$$L = iv \frac{\partial}{\partial x}.$$

Then equation (4.4) becomes:

$$\frac{\partial W}{\partial t} = v \frac{\partial W}{\partial x} + v^2 \tau \frac{\partial^2 W}{\partial x^2}. \tag{4.5}$$

Equation (4.5) describes diffusion (with the diffusion coefficient equal to $v^2 \tau$) with a drift (with the drift velocity equal to *v*).

Similarly to the quantum mechanical case, we assume that the equations of motion in the microscopic random time are usual classical equations. Then their solutions (and, consequently, the trajectories in the phase space of the system) do not change. What is modified (and becomes a random quantity) is the location on the trajectory. Therefore, all conservation laws of classical mechanics hold also for the modified dynamics.

For an illustration, let us continue with the example of the free motion of a particle. The location of the particle is defined by the equality

$$\xi = x^0 + v\theta, \tag{4.6}$$

where $x^0$ is its initial location, $v$ is its velocity, and $\theta$ is the microscopic time. The location $\xi$ is random owing to randomness of the time $\theta$. If $t$ is the mean (macroscopic) time, then the mean location of the particle is equal to

$$x = x^0 + vt ,\qquad (4.7)$$

and the dispersion of the particle's location is

$$D_\xi = v^2 D_\theta ,\qquad (4.8)$$

where $D_\theta$ is the dispersion of the "microscopic" time. As we discussed earlier [in particular, for the Gaussian distribution (3.11)],

$$D_\theta = t\tau .\qquad (4.9)$$

We obtain that the motion (4.6) is a one-dimensional diffusion on the trajectory line (the diffusion coefficient is equal to $v^2\tau$) with a drift along the line, -- in accordance with what has been established earlier using the modified Liouville equation (4.5).

The random microscopic time $\theta$ parametrizes the trajectory in the system's phase space, and the randomness of the location on the classical phase trajectory originates entirely from the randomness of the microscopic time $\theta$.

In the relativistic case, the situation is essentially the same. The trajectory in 4-dimensional space-time is now parametrized by a random proper time of the particle. For example, in the case of free motion the random time $\theta$ is expressed through the random proper time $\varsigma$ by the equality:

$$\theta = u\varsigma ,\qquad (4.10)$$

where $u=(1-v^2/c^2)^{-1/2}$ is the time component of the 4-velocity. Averaging the equality (4.10) leads to the usual relation between the mean macroscopic time $t$ and the mean macroscopic proper time $s$:

$$t = us .\qquad (4.11)$$

The relation between the dispersion $D_\theta$ of the microscopic time $\theta$ and the dispersion $D_\varsigma$ of the random proper time $\varsigma$, as it follows from (4.10), is given by the equality

$$D_\theta = u^2 D_\varsigma .\qquad (4.12)$$

If the random microscopic proper time $\varsigma$ is a generalized Poisson (or Gaussian) process $\varsigma(s)$ in the macroscopic proper time $s$, then its dispersion is equal to $D_\varsigma = s\tau_0$ (where $\tau_0$ is a Lorenz-invariant constant). Therefore (4.12) and (4.11) yield expression (4.9) for the dispersion of the microscopic time $\theta$ with

$$\tau = u\tau_0, \tag{4.13}$$

i.e. the time interval $\tau$ depends on the velocity of the particle, as it was assumed earlier [see equation (2.2)].

In summary, in the relativistic case the 4-dimensional time-space is not random per se. The random ingredient is the interval. Then, similarly to the non-relativistic case, the motion in the space-time – and, consequently, the location on the trajectory – are random.

**5. Absolute time in relativistic and quantum physics**. Farther we summarize the main ideas of the theories with the invariant time parameter, which were put forward in the first half of the XX-th century by Fock [25], Schtueckelberg [26], Feynman [27], Schwinger [28] and further developed during the last decades by many authors – in particular, by Horwitz, Piron [29], Fanchi [30], Garcia-Alvarez [31], Kypriandis [32] and others. The review of this approach can be found in the book [33].

The invariant time parameter will be called "absolute time".

We assume that for a classical system the equations of motion in the absolute time $\zeta$, which is assumed to be random, has the Hamilton form. Let us remind how the evolution in a random time is described. The independent variable in the initial equations is a random "microscopic" time, in this case it is $\zeta$. However, after these equations are solved and the physical quantities are represented as functions of the variable $\zeta$, one should average the results over some probability distribution $P(\zeta|s)$, where $s$ is the average "macroscopic" time, in which the evolution is considered. For a free particle with the mass $m$ the relativistic Hamiltonian is:

$$H = \frac{p_\alpha p^\alpha}{2m} = \frac{p_0^2 - p^2}{2m}, \tag{5.1}$$

where $p_\alpha$ is the 4-momentum, $\vec{p}$ is the three-dimensional momentum vector, $p_0 = \dfrac{E}{c}$ is the energy divided by the speed of light. For a system of non-interacting particles the Hamiltonian is a sum of the terms of the form (5.1). The equation of motion for each particle with such a Hamiltonian can be written directly in the mean "macroscopic" time $s$ in the form[2]:

$$\frac{dx_\alpha}{ds} = \frac{\partial H}{\partial p^\alpha} = \frac{p_\alpha}{m}, \qquad \frac{dp_\alpha}{ds} = 0,$$  (5.2)

i.e. the 4-momentum $p_\alpha$ is a constant.

Here $x_\alpha$ is the 4-vector of the coordinate, $x_0 = ct$ is the time multiplied by the speed of light, and the three remaining components are the coordinates of the particle's position vector. If at the initial moment of time the particle was on-shell, then it will stay on shell as a result of conservation of the 4-momentum, i.e. at any moment of time it will be true that

$$p_\alpha p^\alpha = (mc)^2.$$  (5.3)

The square of the interval (of the proper time) is $(d\bar{s})^2 = \frac{1}{c^2} dx_\alpha dx^\alpha$, and by virtue of (5.2)

$$\frac{d\bar{s}}{ds} = \sqrt{\frac{p_\alpha p^\alpha}{(mc)^2}}.$$  (5.4)

Hence, by virtue of (5.3) $\frac{d\bar{s}}{ds} = 1$, i.e. *the proper time of a particle coincides with the absolute time*. Therefore, any clock synchronized according to Einstein (with the account of the fact that the speed of light is constant) show the absolute time in any coordinate system, in which they are at rest. The conclusion that the proper times and the absolute time coincide remains valid also for a system of non-interacting particles: the proper time of each of them coincides with the absolute time. If the particles interact with each other, then the relation between the proper time of a particle with the absolute time is, generally speaking, not so simple. However, if an interacting particle remains on-shell, then the conclusion that its proper time coincides with the absolute time remains valid by virtue of (5.4). Therefore, randomness in the flow of time has impact only on the intervals between the events, i.e. on the flow of the proper time. In the non-relativistic case, when the velocities are small in comparison with the speed of light, the proper time almost coincides with the Minkowski time, being simultaneously the Galileo–Newton time. Thus, the latter coincides with the absolute time.

---

[2] It is easy to see that for free non-interacting particles the mean values of the physical quantities treated as functions of the variable $s$ satisfy the initial equations with the independent variable $\zeta$ replaced by $s$.

In quantum mechanics, in the framework of the perturbation theory (when the perturbations are small), the proper time method is practically the same as "the absolute time method". Therefore, for low-energy interactions and, thus, on macroscopic (but not cosmological) and atomic scales the absolute time can be said to coincide with the relative relativistic $t$ time.

The relativistic quantum equation of motion for a particle in the absolute time is:

$$ih\frac{\partial \psi}{\partial \zeta} = H\psi \,, \tag{5.5}$$

where $H$ is the Hamiltonian. The mean values of the physical quantities, the evolution of which is calculated according to (5.1) as a function of the random time $\zeta$, should be averaged over some probability density $P\big(\zeta\,|s\big)$ and written as functions of the random time $s$. For scalar particles (bosons with spin 0)[3], $\psi$ is a scalar function (generally, of a complex variable), and the Hamiltonian can be written in the form:

$$H = -\frac{h^2}{m}\partial_\alpha \partial^\alpha \,. \tag{5.6}$$

This Hamiltonian is the quantum version of the Hamiltonian (5.1) (up to a coefficient 1/2), where the components of the momentum operator are given in the coordinate representation $p_\alpha = ih\dfrac{\partial}{\partial x^\alpha}$.

For particles with spin ½ (fermions), $\psi$ is a four-component spinor, and the Hamiltonian can be written in the form[4]:

$$H = ihc\gamma^\alpha \partial_\alpha \,, \tag{5.7}$$

---

[3] Usually the relativistic quantum evolution equation with invariant time parameter is written with the Hamiltonian, which has an extra factor of ½ with respect to the Hamiltonian of equation (5.5), i.e. usually the classical Hamiltonian (5.1) is directly quantized.

[4] This form of the Hamiltonian utilizing the first order derivatives with respect to time and space was used by Feynman. Fock, Schwinger, and later Sakharov considered the Hamiltonian with the second order derivatives.

where $\partial_\alpha = \dfrac{\partial}{\partial x^\alpha}$ are the differentiation operators, $\gamma^\alpha$ are the Dirac matrices. For the Dirac equation written in the form of a second order equation, the Hamiltonian (5.5) should be written in the form:

$$H = -\frac{h^2}{m}\left(\gamma^\alpha \partial_\alpha\right)^2 .$$
(5.7*)

The Klein-Gordon and Dirac equations for the (independent of the absolute time $s$) wave function $\phi$ of the particles, which are on-shell, can be obtained by setting in (5.5)

$$\psi = \exp\left(-\frac{imc^2\zeta}{h}\right)\phi .$$
(5.8)

In order to write the equations of motion for a particle in an electromagnetic field, one should replace, as usual, $p_\alpha = h\partial_\alpha$ with $p_\alpha - \dfrac{e}{c}A_\alpha$ in (5.6), (5.7) and (5.7*) (here $A_\alpha$ is the four-dimensional vector potential of the field).

The wave function $\psi\left(x|\zeta\right)$ allows for an interpretation, which is analogous to the interpretation from conventional non-relativistic quantum mechanics, with the difference that the four-dimensional Minkowski space is assumed instead of usual three-dimensional space. In particular, the quantity $\rho\left(x|\zeta\right) = \left|\psi\left(x|\zeta\right)\right|^2$ is the probability density in the Minkowski space, which can be normalized to 1:

$$\int \rho\left(x|\zeta\right)d^4x = 1 .$$
(5.9)

This normalization holds for any moment of time $\varsigma$, since, as it is easy to show, by virtue of (5.5) the derivative of $\rho\left(x|\zeta\right)$ with respect to time is equal to the four-dimensional divergence:

$$\frac{\partial \rho}{\partial \zeta} = \partial_\alpha J^\alpha ,$$
(5.10)

where

$$J^\alpha = \frac{1}{ih}\left(\psi * \partial^\alpha \psi - \psi \partial^\alpha \psi *\right)$$
(5.11)

for the Hamiltonian (5.6) (the asterix* denotes complex conjugate) and

$$J^\alpha = \psi^+ \gamma^0 \gamma^\alpha \psi$$
(5.12)

for the Hamiltonian (5.7) (the superscript $^+$ denotes, as usual, transposition combined with complex conjugation). If $J^0$ is negative, this can be interpreted, as usual, as presence of an antiparticle with a negative density of charge. For the particles with spin ½ the density $J^0 = \psi^+\psi$ is always non-negative (it follows form (5.12) by virtue of the fact that $\left(\gamma^0\right)^2$ is the identity matrix). It means that equation (5.5) with the Hamiltonian (5.7) always describes particles, but not antiparticles.

An experimental evidence that considering the wave function in the Minkowski space is possible is the observation of interference from two "time slits" [34]. The interpretation of this experiment in the framework of the concept of the wave function evolution in the absolute time is given in [35].

In the literature one can often find the "uncertainty relation" for the mass and the proper time, which reads:

$$\Delta\zeta\Delta m \geq \frac{h}{c^2}.$$

(5.13)

The attempts to justify this relation considering the absolute time and mass as adjoint operators, which are often made in the literature, do not withstand criticism. Indeed, in accordance with the main idea of the discussed approach, the absolute time is a parameter, and not an operator[5]. An interpretation of the relation (5.13) in terms of an equilibrium statistical ensemble of particles, which can leave the mass shell as a result of interactions, is proposed in [36]. In this interpretation $\Delta m$ is the standard deviation from the mass shell, $\Delta\zeta$ is the minimal time required for a particle to escape the minimal macroscopic 4-volume, i.e. the minimal volume enclosing the set of particles[6] (the sample), which is representative for the given ensemble. An equilibrium statistical ensemble of particles, which can leave the mass shell, is constructed in [37] in a regular way as an ensemble with the maximal entropy with a fixed (in addition to all other fixed values) mean value of the mass.

---

[5]Similarly, wrong are the attempts to prove the energy-time uncertainty relation in non-relativistic quantum mechanics by considering energy and time as operators. The energy-time relation in non-relativistic quantum mechanics has a fundamentally different nature.

[6] In [36], as well as in their other works, the authors use the term "ensemble of events", and not "ensemble of particles". Here the "event" is a point in the Minkowski space, which represents the state of a particle. In my opinion, using the term "ensemble of events" in place of "ensemble of particles" is not substantiated, since when the state of a particle is described by a point in the phase space we talk about an ensemble of particles, and not about an "ensemble of phase points".

At present time it is already clear that the "five-dimensional" formalism is as good as the conventional descriptions both in quantum mechanics of particles and in quantum field theory, and sometimes it even appears to be more convenient. However, it is not clear up to now if this formalism, combined with its proper interpretation, could lead to fundamentally new physical results, which cannot be obtained by other means. From the point of view of the "random time" hypothesis, it is the Newton time which is random. The "Newton clock" count the same absolute time in all coordinate systems – the time "ticks" in the same way everywhere (the number of the time ticks in an interval of the "mean" Newton time is random, but it is the same everywhere), while the intervals between the ticks are relative (i.e. depend on the coordinate system). The intervals between the ticks in the proper time are Lorentz invariant.

If absolute time is real and proper times of particles count the absolute time then proper times of all particles are the same. But the direct comparison proper times of two particles is possible only if the particles have the same trajectories because proper times indicated by clocks moving together with particles. In other cases the comparison of clocks moving together with particles may be done with different clocks moving relatively the clock indicated proper time. Therefore, a direct verification of whether the two proper times are "the same" or not is impossible by the definition of the proper time.

However, the fact that it is impossible to compare two proper times directly does not necessarily mean that such a comparison is impossible on principle. One can verify consequences of the hypothesis that they are the same. The key idea is as follows. Consider the system of the equations of motion for two charges, which are written in a relativistically invariant form with the proper times of the particles treated as independent variables. To formulate the problem of making a comparison between the proper times $\xi^1$ and $\xi^2$ of two interacting charges, let us write down their equations of motion in relativistically invariant form. For that we will parametrize the particles' trajectories in the Minkowski space by the proper times $\xi^1$ and $\xi^2$ of the particle treated as independent variables. The equations of motion of two charges can be written as a system of two Lorentz equations, in which the Lorentz force acting on one of the charges from the another one is written via the Liénard–Wiechert potential due to the second charge. The Liénard–Wiechert potentials and the four-dimensional Lorentz force can be expressed via the four-dimensional coordinates and the proper time of the particle, which creates the field. The corresponding expressions can be found in the book [38]. Using the four-dimensional Lorentz force as a function of the coordinates, velocities and accelerations of the charges (see Eq.(245) in [38]) , we can rewrite the system of equations of motion for the charges as equations for 4-coordinates of particles with retarded arguments. Finally we put $\xi^{1} = \xi^{2} = \xi$. If we could prove the existence and uniqueness  of the solution of received differential equations with independent variable $\xi$ then this solution may be compared in principle with observations.

If observations confirm the correctness of the calculated trajectories of particles in Minkowski space, then we can assume the reality of absolute time.

There is no doubt that in the classical (non-quantum) case under consideration, the usual relativistic equations of motion of interacting charges are correct (correspond to possible observations). Therefore, if the trajectories in Minkowski space calculated by the equations with the independent variable $\xi$ coincide with the corresponding trajectories calculated by the usual equations of relativistic dynamics, then there is a possibility that absolute time is real (and there is no need to make special observations).

An indirect argument for the equality of the proper times is the equality of the decay times of identical particles. Horowitz[7] substantiates the identical course of different clocks arguing that this assumption agrees with the observation of the red shift in the gravity field.

Let's briefly discuss how absolute time can be introduced into string theory.

In string theory, for the absolute time one should take the evolution parameter in the proper gauge.

Let us consider a free open bosonic string in the $D$-dimension Minkowski space. Let $\zeta, \sigma$ denote the coordinates on the 2-surface, which represents the path of the string, and $\zeta$ is the time-like evolution parameter, and $\sigma$ $\left(-\pi \leq \sigma \leq \pi\right)$ is the parameter describing the point on the string. The location of the string in the Minkowski space is defined by the coordinates:

$$X^{\mu}\left(\sigma,\zeta\right) \quad \mu = 0, 1, ..., D-1 ,$$

---

[7] See "Time and the Evolution of States in Relativistic Classical and Quantum Mechanics" (arXiv:hep-ph/9606330).

where $\mu = 0$ corresponds to the time coordinate, and all other values of $\mu$ correspond to the space coordinates. Let us assume that all the coordinates $X^{\mu}$ (including the time coordinate), as well as the evolution parameter $\zeta$ has the units of length. We will denote by a dot the derivative with respect to $\zeta$, and by a prime – the derivative with respect to $\sigma$:

$$\dot{X}^{\mu} = \frac{\partial X^{\mu}}{\partial \zeta}, \quad X'^{\mu} = \frac{\partial X^{\mu}}{\partial \sigma}.$$

The boundary conditions for an open string are:

$$X'^{\mu}(-\pi, \zeta) = X'^{\mu}(\pi, \zeta) = 0.$$

We will also use the following shortcut notations for $D$-vectors $X^{\mu}$ and $Y^{\mu}$ in the Minkowski space: $X^{\mu}Y_{\mu} = XY$, and, in particular, $X^{\mu}X_{\mu} = X^2$.

To make the evolution parameter $\zeta$ coincide with the absolute time, let us choose it so that the following conditions are satisfied[8]:

$$\dot{X}X' = 0, \quad \dot{X}^2 + X'^2 = 0. \tag{5.14}$$

In this gauge it is true that:

$$p^{\mu} = -mc\dot{X}^{\mu}, \tag{5.15}$$

where $m$ is the spring mass, and $c$ is the speed of light. Then the Hamiltonian reads[9]:

$$H = -\frac{1}{2\pi mc} \int_{-\pi}^{\pi} \left[ p^2 + (mc)^2 X'^2 \right] d\sigma. \tag{5.16}$$

To make a limiting transition from a string to a point particle one should assume that $X^{\mu}(\sigma, \zeta)$ does not depend on $\sigma$, and set $X'^{\mu} = 0$. Then the Hamiltonian (5.16) takes the form of the Hamiltonian of a point particle (5.1) with the coefficient $\frac{1}{c}$, as it indeed should be, since by the adopted convention the evolution parameter of the string has the units of length.

---

[8] The main notion behind the choice of the gauge is enforcing that the evolution parameter of the string coincides with the absolute time of a point particle under the limiting transition from the former to the latter.

[9] Relations (5.14) – (5.16) are based on the results of Chapter 2 in the book [39] (see also pp. 392 – 404 of the review Marinov M. S., Relativistic strings and dual models of the strong interactions, Sov. Phys. Usp. v. 20, pp. 179–208 (1977).

In conclusion, let us stress once again that in accordance with the hypothesis on randomness in the flow of time, the random time is not the Einstein–Minkowski time, but the absolute time[10]. We note also that the hypothesis on randomness in the flow of time can be applied to the absolute time both in the framework of relativistic one-particle theory and in the framework of field theory.

It was suggested in Appendix C that randomness in the flow of time might be a relic of quantum gravity. An attempt to connect random proper time with the dynamical triangulation model, which is used in the theory of quantum gravity, was made in [40][11].

Random and discrete character of the absolute time results in random and discrete dynamics on the Lorentz and Pseudo-Riemannian manifolds. We note that introducing a discrete random structure on these manifolds, as it was done in the theory of casual sets, and therefore separating "kinematics" from "dynamics", is not required.

**6. Experimental confirmation of the spontaneous decoherence hypothesis and evaluation of the universal time constant.** It is difficult to reveal spontaneous decoherence on a hum noise and to distinguish it experimentally from decoherence due to thermal noise, or other environment-induced factors. Experimental observation of spontaneous decoherence requires strong isolation of the investigated system from the environment.

There are two crucial features that distinguish spontaneous decoherence from the environment-induced decoherence: (i) the time scale at which spontaneous decoherence occurs depends decisively on the energy difference between the states forming the superposition; (ii) if

---

[10] The term "absolute (Newton) time" seems more appropriate to me than frequently used terms "historical time", "evolution parameter" and "invariant time parameter".

[11] In [40] random time is connected with the stochastic time of the stochastic quantization formalism, which is not intrinsically random: there randomness appears in the equations of dynamics as an additional noise. However, the Fokker–Plank equation, which describes the dynamics in the stochastic quantization formalism, also describes a dynamical process with random time, with the mean macroscopic time of the hypothesis on randomness in the flow of time playing the role of the stochastic time of the stochastic quantization formalism. Therefore, the hypothesis on randomness in the flow of time yields the same conclusions as the stochastic quantization formalism used in [40], and hence the results of [40] can be viewed as a possibility to consider the proper time of the studied model of quantum gravity as random time.

The random character of the proper time follows from randomness of time-like geodesics in "statistical geometry" [41].

an experiment is conducted with moving particles, then the spontaneous decoherence time scale depends on their velocities, by virtue of (2.1) and (2.2).

An experimental evaluation of the spontaneous decoherence effect may be provided by direct experiments on interference of two electron beams with slightly different energies, $E_{1,2} = E \pm \Delta E$, as proposed in [42]. Spontaneous decoherence occurs if the time of flight of the electrons from the source to the location where the beams interfere, $t_f$, exceeds the decoherence time $t_{dec}$. A simple calculation, which takes into account (2.1), (2.2), and the relativistic dependence of the electron's energy on its velocity leads to the following condition for spontaneous decoherence:

$$\Delta E > \left(\frac{\pi}{2}\right) h \left(\frac{c}{l\tau_0}\right)^{\frac{1}{2}} \left(1 - \gamma^2\right)^{\frac{1}{2}} \gamma, \quad \gamma = \left(1 - \left(\frac{v}{c}\right)^2\right)^{\frac{1}{2}} = \frac{mc^2}{E},$$  (6.1)

where $m$ is the electron mass, $c$ is the speed of light, and $l$ is the distance between the source of the electrons and the location where the beams interfere.

The successful outcome of the experiment with the electron beams would be that the beats discussed in [42] do not occur as a result of decoherence. This would happen if inequality (6.1) holds. Then the constant $\tau_0$ could be estimated from (6.1).

One way to observe decoherence is by observing interference of two beams, which are split similarly to the Stern-Gerlach experiment[12], given that the beams meet at a point, which is far enough from the place where they split.

Alternatively, one can use Aharonov-Bohm effect[13], where two electron beams interfere. One of the beams passes through a constant electric potential $V$. If the time required for this beam to reach the meeting point is larger than $(\pi h)^2/\tau(eV)^2$, then the interference picture will not appear.

One more way of confirming the existence of spontaneous decoherence and estimating the universal time constant is by observing quantum oscillations. Quantum oscillations occur when an elementary particle is in a coherent superposition of its non-degenerate quantum states, which is exactly the situation when spontaneous decoherence becomes possible. Whether or not decoherence takes place depends on the relationship between the period of oscillations and the decoherence time. Therefore, the very fact that the oscillations can be observed yields an estimate of the universal time interval $\tau$. According to (2.1), the decoherence time is

$$t_{dec} = \frac{\pi^2 h^2}{\tau \Delta E^2} \ .$$

where $\Delta E$ is the energy separation of the involved states. On the other hand, the half-period of the oscillations is equal to

$$t_{os} = \frac{\pi h}{\Delta E}. \tag{6.2}$$

---

[12] The idea of such an experiment was pointed out in: *Wigner E.*, Amer. Jour. Phys., Vol. 31, p. 8.

[13] In the work *Walstad A.*, A Critical Reexamination of the Electrostatic Aharonov-Bohm effect, arXiv:1607.06412 it was shown that purely electrostatic Aharonov-Bohm effect does not exist. However, interference can be observed if a finite vector potential is applied during the experiment.

Therefore,

$$t_{dec} = \frac{(t_{os}^2)}{\tau} \ . \tag{6.3}$$

The necessary condition for the oscillations is $t_{dec} > t_{os}$, and we obtain the estimate:

$$\tau < t_{os}. \tag{6.4}$$

However, in this setting the beam of particles is observed only at one specific location, it is not traced along the whole way from its source. The necessary condition for observing oscillations at that location is that coherence is not destroyed during the flight time $t_f$, i.e. $t_{dec} > t_f$. This requires an estimate stronger than (6.4), namely,

$$\tau < \frac{(t_{os})^2}{t_f}. \tag{6.5}$$

It is obvious that $t_{os}/t_f < 1$. For oscillations of particles with non-relativistic speeds we can set $\Delta E = \Delta mc^2$. For ultra-relativistic case, we have:

$$\Delta E = \frac{(\Delta m)^2 c^4}{E}. \tag{6.6}$$

Equation (6.6) can be easily derived from the relativistic dependence of particle's energy on its momentum.

To this day, the oscillations of $K^0$-, $B$-mesons and neutrinos have been observed. Using the data available from these experiments we obtain the estimates $\tau < 10^{-11}$ sec for mesons and $\tau < 10^{-7}$ sec for neutrino, which, unfortunately, are obviously true and thus not of practical interest.

The best estimate for the universal time constant can be done on the basis of the data for the radiative decay. Equation (3.15) means that any observed lifetime always exceeds the universal time interval, $\tau$. As lifetimes of the order of $10^{-23}$ sec are known, we have $\tau < 10^{-23}$c.

However, were the constant $\tau$ to be even much smaller, say, $\tau \approx 10^{-30}$ sec, then two wave packets having, e.g., an energy difference of the order of $\Delta E \approx 10^{-6}$ erg would lose coherence and the interference would be destroyed at the time interval of the order of $10^{-12}$ sec. Therefore, the states of a macroscopic body cannot produce superpositions, because such superpositions are destroyed spontaneously and very quickly. Note that even for the time interval $t=10^{10}$ years, the standard deviation of the time flow does not exceed $10^{-5}$ sec if $\tau \approx 10^{-30}$ sec.

Clearly, a strict isolation of a system from the outside factors such as thermal noise is necessary for successful observation of spontaneous decoherence. Besides that, the difference of the energy values of the states forming the superposition must not be too small, so that the decoherence time does not exceed the duration of the experiment (as was obviously the case, for example, in the experiments on verifying the nonlinear modification of quantum mechanics, [43], where the superposition of hyperfine levels was investigated: if the energy difference ΔE is such that the frequency ω=ΔE /h = $10^{11}$ sec$^{-1}$, then the decoherence time may be equal to dozens of minutes provided that τ=$10^{-24}$ sec, and even dozens of years provided that τ=$10^{-30}$ sec).

### Addendum

We call a random time modular one if clock ticks $n$ times in a mean (macroscopic) time $t$, where $n$ is a random number, which follows the Poisson distribution with the average $n_{av} = \dfrac{t}{\tau}$, where the τ is a constant, which we call the module of time.

Having familiarized myself with Aharonov's article[14], I understood that the value $c\tau$ is a module in Heisenberg's modular representation suggested by Aharonov. In contrast to Aharonov, where the value of the module is determined by the experimental device (the distance between the slots in the screen), the value τ is a fundamental constant and is determined by the objective properties of space-time. Our Modular representation defines the space-time and its properties. The modular structure introduced by Aharonov is called contextual and environmentally conditioned in [44], where the module is also assumed to be a fundamental constant.

---

[14] "On the Aharonov-Bohm Effect and Why Heisenberg Captures Nonlocality Better than Shredinger", Lecture delivered at the Tonamura First International Symposium on Electron Microscopy and Gauge Fields Tokyo, May 10-12, 2012.

I thank L.P. Pitaevski for his stimulating criticism and M.I. Kaganov for his useful advices. I am especially grateful to L. Accardi for his kind attention and comments on my paper.

Many thanks to P. Chebotarev and V. Yakuba for their invaluable help in preparing my works for publication.

### Appendix A: Some techniques of introducing randomness into dynamics

Let us discuss possible techniques of introducing randomness into dynamics.

To begin with, dynamics of a physical system is considered on the manifold of its states called the "state space". The state space of a system is endowed with some mathematical structure (at least, a topology and the closeness of the states are introduced).

The state space of a system in classical mechanics is a finite-dimensional phase space. The state space of a quantum system described by pure states is the Hilbert space. The state space of a classical system in statistical physics is the space of probability measures on the phase space (the ensemble of non-negative integrable probability density functions normalized to 1). The state space of a quantum system in a mixed state is the ensemble of the density operators, i.e. self-adjoint operators with traces equal to 1.

Let us define the concept of a trajectory in the state space. A trajectory is an image of the real non-negative axis $[0, \infty)$ ("time"), continuously mapped onto the state space. A trajectory is determined by the evolution equation and the initial state. Evolution equations are usually some differential equations, in which time is an independent variable. The evolution equations of a system in classical mechanics are ordinary differential Hamilton equations. The evolution equation of a quantum system in its pure state is the Shrödinger equation (or an equivalent description). The evolution equation of a classical system in statistical physics is the Liouville equation. The evolution equation of a quantum system in a mixed states is the quantum Liouville equation, known also as J. von Neumann's equation. In all these cases, the trajectory is determined by the unique solution of the Cauchy problem for the evolution equations with some initial conditions.

Random time can be introduced naturally by two different techniques. The first technique introduces randomness into the evolution equations. It is possible, for example, to use stochastic differential equations instead of differential equations to describe evolution. It would mean randomness of the trajectories. The second technique allows time to be a stochastic parameter, and preserves the deterministic character of the evolution equations. Then the trajectories are deterministic, but the positions on the trajectories become random.

Comparing these two techniques we can see that the first technique is much more radical than the second one. The second technique preserves all basic principles of original theories, including the evolution equations (in the microscopic time). The evolution is non-unitary only in the macroscopic time. In contrast, the first technique can give rise to difficulties from the very outset. For instance, in the quantum case, using stochastic differential equations would violate the unitarity of evolution, and, as a result, the principles of quantization would need revision. For the second technique, all the conservation laws of the original theory remain valid because the trajectories are not changed, while the first technique may lead to a violation of the conservation laws of the original theory. It was shown in [45] that, indeed, stochastic differential equations are the cause of violation of the energy conservation in the collapse model [4].

This comparison argues in favor of the second technique.

The heuristic ideas on the role of random time in spontaneous decoherence are consistent only with the second technique. Indeed, randomization of the phase relations between stationary states is a meaningful explanation of spontaneous decoherence only if the modified theory preserves the concept of stationary states and the linear increase (with the increase of time) of the phases of their wave functions. This requires the conventional notion of quantum trajectories in the Hilbert space. The randomness of the phase differences is attributed to the randomness of one parameter of the evolution equations – that is, time.

## Appendix B: Non-decreasing of entropy

The time derivative of the entropy, $S = -Tr\left(R \ln R\right) = -\sum_k \lambda_k \ln \lambda_k$, can be written as

$$\frac{dS}{dt} = -\sum_k \left(\frac{d\lambda_k}{dt}\right)\ln\left(\frac{1}{\lambda_k}\right),$$ (B.1)

where $\lambda_k$ are the eigenvalues of the matrix $R$. Here we took into account that $\dfrac{d\sum_k \lambda_k}{dt} = 0$ .

For each given moment $t$, let us consider a representation, in which the density matrix $R(t)$ is diagonal. This representation and the eigenvalues $\lambda_k$ depend on $t$. From the characteristic equation

$$Det\left[R(t) - \lambda I\right] = 0$$ (B.2)

(where $I$ is the unit matrix) we can conclude that at the moment $t$ the eigenvalues are equal to the diagonal elements of the matrix $R(t)$:

- $$\lambda_k = R_{kk}(t).$$

(B.3)

However, the derivatives $l_k = \dfrac{d\lambda_k}{dt}$ are not equal to the derivatives of the corresponding diagonal matrix elements, because the matrix $R$ is not necessarily diagonal at later moments of time in that representation, in which it was diagonal at the moment $t$. Let us write down the equation for the moment $t+dt$:

$$Det\left[R(t+dt) - \lambda(t+dt)I\right] = 0$$

and represent it in the form:

$$Det\left[R(t) + R'(t)dt - (\lambda_k + l_k dt)I\right] = 0,$$ (B.4)

where $R'(t)$ is the time derivative of the matrix $R(t)$.

The derivatives $l_k$ can be found from (B.4) with the account of (B.3) by equating to zero the coefficient at the lowest degree of $dt$ in the determinant of expression (B.4). If there is no degeneracy (i.e. when $\lambda_k$ is not equal to any other eigenvalue), then the lowest power of the differential $dt$ is 1, and $l_k = R'_{kk}(t)$. However, if there is $r$-fold degeneracy, then the lowest power of the differential $dt$ is $r$. Assume that first $r$ eigenvalues $\lambda_k = R_{kk}(t)$, $k=1,...,r$ are equal to each other. Then their derivatives $l_k = \dfrac{d\lambda_k}{dt}$, $k=1,...r$, can be found as the roots of the characteristic equation of order $r$:

$$Det\left[\,^{r}R'(t) - lI\right] = 0,$$

(B.5)

where $^{r}R'(t)$ is the time derivative of the $r \times r$ submatrix of the matrix $R(t)$. From (B.5) we conclude that the sum of the derivatives of the eigenvalues is equal to the trace of the matrix $^{r}R'(t)$, i.e.:

$$\sum_{k=1}^{k=r} l_k = \sum_{k=1}^{k=r} R'_{kk}(t).$$

(B.6)

From (B.1), (B.3), and (B.6) we have the following expression for the entropy, which is valid both for degenerate and non-degenerate cases:

$$\frac{dS}{dt} = -\sum_k R'_{kk} \ln\left(\frac{1}{R_{kk}}\right).$$

(B.7)

Thus, the only remaining task is to determine the derivatives of the diagonal elements of the matrix $R(t)$ in the representation where it is diagonal.

In doing this, let us consider separately the models in which random time follows the generalized Poisson and Gaussian distributions.

*1. Random time with the generalized Poisson distribution.* The evolution equation for the random time obeying the generalized Poisson distribution is equation (3.9), which has the form:

$$\frac{dR}{dt} = \left(\frac{1}{\tau}\right)\left[\int \exp\left(\frac{-iH\xi\tau}{h}\right)R(t)\exp\left(\frac{iH\xi\tau}{h}\right)p(\xi)d\xi - R(t)\right].$$

(B.8)

We introduce the notation

$$Q = \exp\left(\frac{-iH\xi\tau}{h}\right),$$

(B.9)

and $Q^+$ for its conjugate. Taking into account that $\int p(\xi)d\xi = 1$, we have from (B.8):

$$\frac{dR(t)}{dt} = \left(\frac{1}{\tau}\right)\int\left[QR(t)Q^+ - R(t)\right]p(\xi)d\xi.$$

(B.10)

In the representation where the matrix $R(t)$ is diagonal, we have from (B.10) for the derivatives $R'_{kk}$ of the diagonal elements:

$$R'_{kk} = \left(\frac{1}{\tau}\right)\int\left[\sum_j |Q_{kj}|^2 R_{jj}(t) - R_{kk}(t)\right]p(\xi)d\xi.$$

(B.11)

Applying equation (B.7) and reverting to the notation $\lambda_k = R_{kk}(t)$, we have:

$$\frac{dS}{dt} = \left(\frac{1}{\tau}\right)\int\left[\sum_{k,j}A_{kj}\lambda_j\,\ln\left(\frac{1}{\lambda_k}\right) - \sum_k\lambda_k\,\ln\left(\frac{1}{\lambda_k}\right)\right]p(\xi)d\xi\,, \tag{B.12}$$

where $A_{kj} = \left|Q_{kj}\right|^2$. By virtue of unitarity of the matrix $Q$ (which is, of course, unitary in all representations, including the representation where the density matrix $R(t)$ is diagonal), we have:

$$\sum_j A_{kj} = \sum_k A_{kj} = 1, \quad A_{kj} \geq 0\,. \tag{B.13}$$

Let us prove that the integrand in equation (B.12) is non-negative. This assertion, which completes the demonstration of non-decreasing of the entropy for the generalized Poisson time, is a direct consequence of the following lemma.

*Lemma.* Let the sequence of positive numbers $x_1,...,x_k...$ be non-increasing, and the sequence of positive numbers $y_1,...,y_k...$ be non-decreasing. Let the conditions (B.13) be fulfilled for the quantities $A_{kj}$. Then the following inequality holds:

$$\sum_{k,j}A_{kj}x_j\,y_k \geq \sum_k x_k\,y_k\,. \tag{B.14}$$

Indeed, to prove non-negativity of the integrand in equation (B.12), we can apply this lemma to the integrand, with $x_k = \lambda_k$, $y_k = \ln\left(\frac{1}{\lambda_k}\right)$ and with the eigenvalues enumerated in the non-increasing order.

*Lemma's proof.* Let us find the minimum of the left-hand side of inequality (B.14) with respect to the variables $A_{kj}$, under the restriction (B.13). In this problem of linear programming the minimum is reached when each row and each column of the matrix $A$ contains exactly one element equal to 1, and all other elements are equal to 0. It means that

$$\sum_{kj}A_{kj}x_j\,y_k \geq \sum_k x_k z_k\,, \tag{B.15}$$

where $z_1,..., z_k...$ are elements of the sequence $y_1,...,y_k$ renumbered in a certain way. Let us consider the difference $\sum_k x_k z_k - \sum_k x_k y_k = \Sigma_k x_k u_k$, where $u_k = z_k - y_k$. It is easy to prove using mathematical induction that the sum of the sequence $u_1, ... u_k$ is non-negative for any $k$, since the sequence $y_1,...,y_k...$ is non-decreasing, and the sequence $z_1,..., z_k...$ consists of the same elements as the sequence $y_1,...,y_k$, though numbered in a different order. The sum of all $u_k$ is obviously equal to 0, because the sum of all $z_k$ is equal to the sum of all $y_k$. Thus, for any $l$ we have:

$$u_1 + ... + u_l \geq 0, \sum_k u_k = 0. \tag{B.16}$$

where the sum runs over all $k$. The Abel identity (which is the finite differences analogue of the expression for integration by parts) leads to the expression

$$\sum_k x_k u_k = \sum_l (x_l - x_{l+1})(u_1 + ... + u_l) + x_{\min} \sum_k u_k, \tag{B.17}$$

where $x_{min}$ is the last term of the non-increasing sequence $x_1,...,x_k...$. From (B.17) and (B.16) we can conclude that $\sum_k x_k u_k \geq 0$. This inequality together with equation (B.15) leads to inequality (B.14). The lemma is proved.

*2. Random time with the Gaussian distribution.* The evolution equation for the Gaussian random time is equation (3.10):

$$ih \frac{dR(t)}{dt} = [H,R] + \left(\frac{\tau}{ih}\right)[H,[H,R]]. \tag{B.18}$$

In the representation where the matrix $R(t)$ is diagonal, and taking into account that the matrix $H$ is Hermitian (which is true in any representation), we can write the equation for the time derivatives of the diagonal elements:

$$R'_{kk}(t) = \left(\frac{2\tau}{h^2}\right)\left[\sum_j |H_{kj}|^2 R_{jj}(t) - R_{kk}(t)\right]. \tag{B.19}$$

Applying equation (B.7) and using the notation $\lambda_k = R_{kk}(t)$, we get after simmetrizing over the indices j and k:

$$\frac{dS}{dt} = \left(\frac{\tau}{h^2}\right)\sum_{k,j} |H_{kj}|^2 (\lambda_k - \lambda_j) \ln\left(\frac{\lambda_k}{\lambda_j}\right) \geq 0. \tag{B.20}$$

Non-decreasing of entropy is proved. It is easy to prove that the right-hand sides of the equalities in (B.12) and (B.20) are strictly positive if the energy spectrum is non-degenerate and the density matrix is non-diagonal.

**Appendix C: The hypothesis on randomness in the flow of time in the context of quantum gravity.**

First of all, it is important to stress that the hypothesis on randomness in the flow of time deals with the time intervals on macroscopic and atomic time scales, and not on the scale of the Planck time. Therefore, quantum gravity is only tangentially related to the hypothesis on randomness in the flow of time, and the discussion of this hypothesis should be hold in the framework of the semi-classical approximation of quantum gravity.

The main feature of stationary random processes with independent increments, which are the keystone of the hypothesis on randomness in the flow of time, is the proportionality of the standard deviation of the random time to $t^{1/2}$ [20]. It is interesting to note that, as E. Wigner showed, the same proportionality holds for the uncertainty of the quantum measurement of time done by a clock with a finite mass. However, later Wigner's arguments were modified to take gravity into account [46], and the measurement error was found to be proportional to $t^{1/3}$. To evaluate the significance of this discrepancy, we will reproduce and analyse Wigner's arguments.

Following E. Wigner, let us consider a measurement of a distance $l$ between a clock and a mirror, provided that the clock has mass $m$. The light source is co-located with the clock, so that the signal goes from the clock to the mirror and comes back. The quantum uncertainty principle for the coordinate and momentum states that an uncertainty of the clock coordinate, $\delta x$, leads to the uncertainty of its velocity of the order of $\dfrac{h}{m\delta x}$. Therefore, the uncertainty of the coordinate of the clock increases to $\dfrac{2lh}{mc\delta x}$ within the travel time of light, $\dfrac{2l}{c}$. The total uncertainty of the coordinate of the clock, and thereby the error of the distance measurement $\delta l$, becomes equal to:

$$\delta l = \delta x + \frac{2lh}{mc\delta x}. \tag{C.1}$$

This expression has a minimum at certain $\delta x$, and its minimal value is equal to

$$\delta l_{quant} = \left( \frac{hl}{mc} \right)^{\frac{1}{2}}. \tag{C.2}$$

For the corresponding time interval $t = \left( \dfrac{l}{c} \right)$ we have:

$$\sigma_{quant} \approx \left( \frac{ht}{mc^2} \right)^{\frac{1}{2}}. \tag{C.3}$$

It was noted in [47] that gravitation brings another uncertainty into the distance measurement. The condition that the size of the clock must exceed its Schwarzschild radius leads to an additional error in the measurement of the distance:

$$\delta l_{grav} \approx \frac{Gm}{c^2}. \tag{C.4}$$

Total uncertainty, quantum plus gravitational, is equal to:

$$\delta l \approx (hl/mc)^{1/2} + Gm/c^2 . \tag{C.5}$$

Note that (C.5) has a minimum at a certain $m$. At the minimum, we obtain the following estimate for the quantum-gravitational uncertainty:

$$\delta l_{q-gr} \approx \lambda \left( \frac{l}{\lambda} \right)^{\frac{1}{3}}, \tag{C.6}$$

where $\lambda = \left( \frac{hG}{c^3} \right)^{\frac{1}{2}}$ is the Planck length. The authors of [47] showed that there is another way to derive equation (C.6), namely, through the holographic principle of quantum gravity. This principle asserts, roughly speaking, that the degrees of freedom contained in a volume can be "encoded" onto a surface bounding this volume.

Assuming that time intervals are equal to the corresponding distances divided by the speed of light $c$, we get from (C.6) the equation for the uncertainty of the time measurements:

$$\sigma_{q-gr} = \eta \left( \frac{t}{\eta} \right)^{\frac{1}{3}}, \tag{C.7}$$

where $\eta = \left( \frac{hG}{c^5} \right)^{\frac{1}{2}}$ is the Planck time interval. In this line of arguments we followed [47], where it was taken for granted that the ratio of the travelled distance to the travel time for light is a constant, $\frac{l}{t} = c$, though it is not generally true in quantum gravity (see [48] for the analysis).

The fact that the quantum-gravitational fluctuations presumably scale as $t^{\frac{1}{3}}$, and not as $t^{\frac{1}{2}}$ (as it should be according to the hypothesis on randomness in the flow of time), means that, as it was noted in [47], the quantum gravity fluctuations of the space-time distances are not a random process with independent increments. However, let us discuss which space- and time scales require us to take gravitation into account, and equation (C.7) is reasonable. When the measured distance increases, the first term in (C.5) increases, while the second term remains constant. Therefore, the relative role of gravitation decreases with the increase of the measured distances, since the role of purely quantum fluctuations goes up. The ratio of the second term in (C.4) to the first is equal to (after replacing the space distance by the corresponding time interval):

$$\left(\frac{m}{\mu}\right)^2 \left(\frac{h}{tmc^2}\right)^{\frac{1}{2}},\tag{C.8}$$

where $\mu \approx 10^{-5}$ g is the Planck mass. If the mass of the clock is not huge, and time intervals have macroscopic or atomic scale, then the magnitude of the expression (C.8) is very small, and in this case it is possible to ignore the gravitational uncertainty.

To understand when equation (C.7) is reasonable, let us focus on the value of $m$, which minimizes (C.5) and therefore leads to (C.6). This value depends on the time interval, so that for equation (C.7) to be valid for some $t$, it is necessary that the mass $m$ has a specific value. It turns out that for macroscopic and atomic time scales the value of the clock mass is unreasonably large. Indeed, the value of $m$ which minimizes expression (C.5) is equal to:

$$m \approx \mu \left(\frac{\varepsilon t}{h}\right)^{\frac{2}{3}} = \mu \left(\frac{t}{\eta}\right)^{\frac{2}{3}},\tag{C.9}$$

where $\mu \approx 10^{-5}$ g is the Planck mass, $\varepsilon \approx 10^{16}$ erg is the Planck energy, $\eta \approx 10^{-44}$ sec is the Planck time interval. For example, for equation (C.7) to be valid for $t \approx 1$ sec, the mass must have the value of $m \approx 10^{24}$ g. Thus, it is unlikely that equation (C.7) holds for the time scales relevant for the hypothesis on randomness in the flow of time.

As to the alternative (through the holographic principle) route to equation (C.6) derived in [47], for large time scales the formulation of the holographic principle may differ from the "naïve" version used in [47]: the "screen" onto which the degrees of freedom from the volume can be "encoded" is not necessarily the border of the volume (see for example [49]).

Furthermore, let us discuss possible relation of the hypothesis on randomness in the flow of time to the semi-classical approximation of quantum gravity. Randomness and/or non-unitarity in the space-time description can appear in two contexts: (i) When the continuum of the quasi-classical states, the "weave" [50] (see also [51]) is constructed on a discrete background. This can be done by introducing stochastic geometry [52] or random dynamics of the "casual sets" [55]; (ii) From the description of quantum evolution (see, for example, a general discussion in [54], and a competing model in [55], which considers also non-unitary models). In any case, the randomness of the time intervals in the hypothesis on randomness in the flow of time and non-unitary evolution do not contradict to semi-classical approximation of quantum gravity. Note that the "black-hole information paradox" does not apply to the situation when spontaneous decoherence really takes place, and unitarity is violated.

We established that the standard deviations for randomness in the flow of time and for the quantum uncertainty in the measurement of time by a clock of finite mass have the same time dependence, $t^{1/2}$. If it is not a pure coincidence, then by virtue of (C.3) we can put forward the assumption of existence of the universal clock with a finite mass $m$, which is related to the time constant $\tau$ as

$$\tau = h / mc^2. \qquad (C.10)$$

Such an assumption agrees with the idea of a continuous medium (a fluid), which plays the role of the reference system and is composed of particles-clocks [56].

Finally, let us discuss a possible source of time randomness using the ideas of Refs. [57], [58]. Based on the connection of the surface of a black hole with its entropy, and on the thermodynamic properties of the Rindler horizon of events, the authors of [57] proposed that each two-dimensional space-like surface contains the states of a "space-time foam". This space-time foam is identified in [58] with a gas of "microscopic black holes". Extending this idea, it is possible to assume that the whole space-time is filled with the space-time foam, and randomness in the flow of time is a manifestation of a kind of the Brownian motion of a particle in its proper time caused by "collisions" with the space-time foam. From this point of view, the time constant $\tau$ is connected with the diffusion coefficient, which is determined by the parameters of the space-time foam. However, this analogy with the Brownian motion should not be taken too literally.

Firstly, the random process of the time flow takes place only in the proper time of the system, and it does not deflect the system from its trajectory (see Appendix A). Secondly, this random process is universal for all material systems, and therefore it does not depend on the parameters of the system, for example, on the particle's mass.

The stochastic nature of the space-time, appearing as quantum fluctuations, which imprint on the classical space-time, is considered in the model of dynamical triangulations (see [59] and references therein). From the point of view of the thermal hypothesis of time, [60], randomness in the flow of time can be a manifestation of temperature fluctuations of the space-time foam.

I conclude this Appendix with a speculative attempt at connecting the time constant $\tau_0$ with the cosmological constant $\Lambda$.

The two scales related to the cosmological constant $\Lambda$, the space scale $\Lambda^{-1/2}$ and the time scale $\left[(c\Lambda)^{1/2}\right]^{-1}$, are connected, respectively, to the size of the universe and the age of the universe. Therefore, at the first glance, the cosmological constant cannot have any relation to the microscopic scales of the quantum fields' processes, including quantum gravity. A number of attempts at connecting the cosmological constant to the vacuum energy came across the absurdly huge discrepancy between the energy density for quantum fields, on the one hand, and the small observed value of the cosmological constant, on the other hand. The energy density, $\rho_\Lambda$, related to the cosmological constant is equal to:

$$\rho_\Lambda = \Lambda c^4 / 8\pi G. \qquad (C.11)$$

For experimentally available values of $\Lambda$, the density $\rho_\Lambda$ is less than Plank's density $\rho_{Pl} = c^5 / hG^2$ by 120 orders of magnitude.

Other known approaches also failed to provide a reasonable agreement (see, e.g., a review in [60]). This systematic failure suggests that, from the point of view of quantum field theory, the natural value for the cosmological constant is zero [61], while its deviation from zero should be attributed to external perturbations of vacuum by matter.

Some works even claimed that the cosmological constant is not related to quantum theory in any way (see [62] and references therein). Below I consider two hypotheses, which connect the cosmological constant to the quantum theory of gravity and define the space-time scales of relevant processes.

*The first hypothesis* connects $\Lambda$ with the Hubble constant $H$ (and therefore with the age of the Universe):

$$\Lambda = \left(H/c\right)^2 . \qquad (C.11)$$

This hypothesis is based on the Friedman equations, and is confirmed by experimentally observed values of $\Lambda$ and $H$. Denoting the age of the Universe by $\Theta \approx 1/H$, we have from (C.11):

$$\Theta = 1/c\Lambda^{1/2} . \qquad (C.12)$$

*The second hypothesis* determines the mass scale, $m_\Lambda$, (and the corresponding energy scale, $E_\Lambda = m_\Lambda c^2$ ), that gives the density of the gravitational energy equal to $\rho_\Lambda$ (see below). These scales are defined so that

$$E_\Lambda = m_\Lambda c^2 = h/\Theta . \qquad (C.13)$$

The scale $m_\Lambda$ was considered in [61] as the minimal possible mass. It is useful to note that the temperature

$$T_\Lambda = E_\Lambda/k = hc\Lambda^{1/2}/k$$

($k$ is the Boltzmann constant) is equal to the temperature of de Sitter's horizon. We determine, following the arguments of [62], the space scale, $l$, providing the energy density of gravitation for the mass, $m_\Lambda$, equal to $\rho_\Lambda$ . To do it we equate $\rho_\Lambda$ to the energy density of the gravitational interaction of masses $\sim m_\Lambda$ separated by a distance $\sim l$ in a volume $l^3$. We have:

$$G\left(m_\Lambda\right)^2/l^4 = \Lambda c^4/8\pi G .$$

After eliminating the numerical factors and taking into account (C.12) and (C.13), we get:

$$l = \lambda = \left(hG/c^3\right)^{1/2} ,$$

i.e. the energy density determined by the cosmological constant is equal to the energy density of the gravitational energy at the Planck length scale, which is the scale of quantum gravity.

On the contrary, we can treat the mass $m_\Lambda$ as the mass of some extended entity. Then we can find the corresponding space scale, $r_\Lambda$, for which the energy density of this entity is equal to $\rho_\Lambda$. From the equality

$$E_\Lambda/\left(r_\Lambda\right)^3 = \rho_\Lambda$$

and equations (C.10), (C.12), (C.13), after eliminating the numerical factors we get

$$r_\Lambda = \lambda^{2/3} \Lambda^{-1/6},$$

and rewrite it in the form:

$$r_\Lambda = \lambda \left[ \lambda^2 \Lambda \right]^{-1/6}. \tag{C.14}$$

Since $\lambda^2 \Lambda \approx 10^{-120}$, the scale $r_\Lambda$ exceeds the Plank scale by approximately 20 orders of magnitude. The authors of [61] associated the scale $r_\Lambda$ with a classical electron's radius, by the reason of their coinciding orders of magnitude.

However, it is more reasonable to identify the corresponding time scale, $r_\Lambda/c$, with the constant $\tau_0$, which determines randomness in the flow of time. This estimate does not contradict to our previous estimates made on the basis of the particles' decays data. It is possible to think that, using the terminology of paper [63], the Plank scale belongs to the realm of the discrete (combinatorial) theory, while the scale $r_\Lambda$ belongs to the "mesoscopic", semiclassical realm, where the discreteness is averaged and a transition to the continuum is accomplished, however, some deviations from the classical picture can still take place. Randomness in the flow of time can be one of such deviations. The randomness of relativistic intervals is the consequence of randomness of time-like geodesic curves in the theories of quantum gravity, which are based on the "statistical geometry" [41].

## Appendix D: Modification of the generalized Poisson time

In order to exclude very small time intervals and avoid ultraviolet divergence, one should modify the generalized Poisson distribution (3.1) in two aspects. Firstly, it is necessary to eliminate the $\delta$-function in this distribution and, secondly, to introduce the distribution guaranteeing small probabilities of small intervals.

*1. Elimination of the $\delta$-function.* In the main text, we considered a usual random process with independent increments with the starting point being zero, i.e. the random time, $\theta$, is the sum of random variables, $\Delta^n$, n=0, 1,…, with $\Delta^0 =0$:

$$\theta = \Sigma\, \Delta^n.$$

This aspect is reflected in equation (3.1), where the summation starts from n=0, and the n-fold resultant for n=0 by convention is equal to the $\delta$-function. To eliminate the $\delta$-function, we can set $\Delta^0$ to be some random quantity[15] instead of 0. In fact, this random quantity $\Delta^0$ should not necessarily have the same probability distribution as $\Delta^n$ for n≠0. Denoting by $p^0$ and $p$ the probability densities for $\Delta^0$ and $\Delta^n$, n≠0, respectively, we can write down the modified distribution $P^*(\theta, t)$ in the form:

$$P^*(\theta,t) = P(\theta,t) - exp(-t/\tau)\delta(\theta) + exp(l/\tau)(-t/\tau)p^0(\theta/\tau), \qquad (D.1)$$

where $P(\theta, t)$ is given by equation (3.1). The characteristic function of the probability density $P^*(\theta, t)$ is equal to

$$\int P^*(\theta,t)exp(i\lambda\theta)d\theta = exp[(\varphi(\lambda\tau)-1)(t/\tau)] + exp(-t/\tau)(\varphi^0(\lambda\tau)-1) - exp(-t/\tau),$$
$$\qquad (D.2)$$

where φ and $\varphi^0$ are the characteristic functions of $p$ and $p^0$. The quantum dynamics discussed in the main text slightly changes after this transformation, however, for a sufficiently large macroscopic time $t$, equation (D.2) tends asymptotically to the expression for the conventional generalized Poisson process [see (3.3)] -- the memory about the initial time interval is erased, and the description of decoherence from the main text is restored.

*2. Distributions with small probabilities of small intervals.* Among the probability distributions that guarantee a small probability of small intervals there are the so-called gamma distributions, which have the form:

$$p(\xi) = (\xi^{-1}/\Gamma(s))exp(-\xi). \qquad (D.3)$$

The probability density $p(\xi)$ vanishes at ξ=0, and is maximum at ξ=s−1 if s>1. The characteristic function of $p(\xi)$ is equal to

$$\varphi(\lambda) = \int p(\xi)exp(i\lambda\xi)d\xi = 1/(1-i\lambda)^s. \qquad (D.4)$$

The Gamma distribution with *s=1* is the usual Poisson distribution considered earlier. We discuss below the case with *s=2*.The characteristic function in this case is:

$$\varphi(\lambda) = (1 - \lambda^2 + 2i\lambda)/(1 + \lambda^2)^2. \qquad (D.5)$$

We will call the microscopic time with the probability distributions (D.3) with *s=2* as the "*gamma time*". Using equations (3.3) and (D.5) we have for the gamma time*:*

---

[15] For the random modular time (see addendum below) $\Delta^0$ is not a random quantity, but it is equal to the module of the proper time. The modular time was forestalled by G.J. Milburn in [16].

$$\int P(\theta,t)\exp(i\lambda\theta)d\theta = \exp\left\{\left[\left(2i\lambda\theta - 3\lambda^2 t\tau - \lambda^4 t\tau^2\right)\middle/\left(1+\lambda^2\tau^2\right)^2\right]\right\}. \qquad (D.6)$$

All main features of quantum evolution in the gamma time are qualitatively the same as in the Poissonian time and in the Gaussian time. Using equation (D.6), we get for the density matrix in the energy representation:

$$R_{kl}(t) = R_{kl}(0)\exp\left\{\left[1+\left(E_k-E_l\right)^2\tau^2/h^2\right]^{-2}\left[2i\left(E_k-E_l\right)t/h - 3\left(E_k-E_l\right)^2 t\tau/h^2 - \right.\right.$$

$$\left.\left. - \left(E_k-E_l\right)^4 t\tau^3/h^4\right]\right\}. \qquad (D.7)$$

The density matrix becomes diagonal when $t\to\infty$, so that the entropy does not decrease (and it increases if there is no degeneracy).

### Appendix E: Evolution of wave packets in random *Gaussian* time

Let the initial (at the moment $\theta=0$) state of a free particle be a Gaussian wave packet with a width $\delta x$:

$$\psi(x,0) = \left(1/2\pi\delta x^2\right)^{1/4}\exp\left[-x^2/4\delta x^2\right].$$

Consider the evolution of the probability density of its coordinate $\Pi(x,\theta)=|\psi(x,\theta)^2|$ in the microscopic time $\theta$ (provided that $\theta$ is the Gaussian time). We will switch to the macroscopic time $t$ later after averaging the resulting equations.

The evolution of the wave packet in the momentum representation is

$$\varphi(p,\theta) = \left[\frac{1}{\sqrt[4]{2\pi\delta p^2}}\right]\exp\left[-\frac{p^2}{4\delta p^2} - i\frac{p^2\theta}{2mh}\right],$$

where $\delta p = \dfrac{h}{2\delta x}$. Coming back to the coordinate representation, we have for the coordinate probability density:

$$\Pi(x,\theta) = \frac{1}{2\pi h}\iint dp_1 dp_2 \frac{1}{\sqrt{2\pi\delta p^2}}\exp\left[-i\frac{p_2 x}{h} - \frac{p_2^2}{4\delta p^2} - i\frac{p_2^2\theta}{2mh}\right]\exp\left[+i\frac{p_1 x}{h} - \frac{p_1^2}{4\delta p^2} + i\frac{p_1^2\theta}{2mh}\right].$$

Let us average $\Pi(x,\theta)$ over $\theta$, using the expression for the Gaussian time:

$$\int P(\theta;t)\exp(i\lambda\theta)\,d\theta = \exp\left(-i\lambda t - \lambda^2 t\tau\right).$$

For the averaged density $\Pi*(x,t) = \int P(\theta;t)\Pi(x,\theta)d\theta$ we have:

$$\Pi*(x,t) = \frac{1}{2\pi h} \iint dp_1 dp_2 \ \frac{1}{\sqrt{2\pi \delta p^2}} \exp\left[ -i \frac{(p_2 - p_1)x}{h} - i\left(\frac{p_1^2}{2m} - \frac{p_2^2}{2m}\right) \frac{t}{h} - \frac{p_2^2}{\delta p^2} - \frac{p_1^2}{\delta p^2} - \left(\frac{p_1^2}{2m} - \frac{p_2^2}{2m}\right)^2 \frac{t\tau}{h^2} \right]$$

.

Let us set

$$t = \frac{h^2}{\tau \left(\dfrac{\Delta p^2}{2m}\right)^2} \ , \tag{E.1}$$

where $\Delta p = \dfrac{h}{l}$, and $l$ is some arbitrary chosen length. We change the variables in the integral for $\Pi*(x,t)$ as follows:

$$p = \frac{p_2 + p_1}{\Delta p} \ , \qquad q = \frac{p_2 - p_1}{\Delta p} \ ,$$

and get:

$$\Pi*(x,t) = (2\pi)^{-\frac{3}{2}} \frac{\delta x}{l^2} \iint dp\,dq \exp\left[ i\frac{qx}{2l} - i4\,pq\left(\frac{2mh}{\tau \Delta p^2}\right) - \frac{(l)^2}{4\delta x^2}\left(p^2 + q^2\right) - p^2 q^2 \right]$$

or

$$\Pi*(x,t) = (2\pi)^{-\frac{3}{2}} \frac{\delta x}{l^2} \iint dp\,dq \cos q\left(\frac{x}{2l} - \frac{8\,pmh}{\tau \Delta p^2}\right) \exp\left[ -\frac{(l)^2}{4\delta x^2}\left(p^2 + q^2\right) - p^2 q^2 \right].$$

After the integration over $q$ we have:

$$\Pi*(x,t) = \frac{1}{2\pi} \frac{\delta x}{l^2} \int dp\ \frac{1}{\sqrt{\left(p^2 + \dfrac{l^2}{4\delta x^2}\right)}} \exp\left[ -\frac{(l)^2}{4\delta x^2}\,p^2 - \frac{\left(\dfrac{x}{2l} - \dfrac{8\,pmh}{\tau \Delta p^2}\right)^2}{4\left(p^2 + \dfrac{l^2}{4\delta x^2}\right)} \right],$$

which brings us to the inequality:

$$\Pi*(x,t) \le \frac{1}{2\pi} \frac{\delta x}{l^2} \int dp\ \frac{1}{\sqrt{\left(\dfrac{l}{2\delta x}\right)^2}} \exp\left[ -\left(\frac{l}{2\delta x}\right)^2 p^2 \right].$$

After the integration over $p$ we have:

$$\Pi*(x,t) \le \frac{2}{\sqrt{\pi}} \frac{\delta x^3}{l^4} . \tag{E.2}$$

Finally, we obtain from (E.1) and (E.2):

$$\Pi *(x,t) \leq \frac{8m^2(\delta x)^3}{\sqrt{\pi}\,\hbar^2}\frac{1}{t} \ . \tag{E.3}$$

This estimate holds also for the wave packets centered at any arbitrary point $x=x_0$, not only at $x=0$. Therefore, the same rate of spreading, $t^{-1}$, applies for any superposition of Gaussian wave packets.

In the conventional quantum mechanics, where the time is deterministic (i.e. $\theta = t$), the probability density is proportional to $\dfrac{1}{\sqrt{\delta x^2 + \left(\dfrac{\delta p}{m}\right)^2 t^2}}$, and therefore the following estimate is true:

$$\Pi_{ort}(x,t) \approx \frac{2m\delta x}{\hbar}\frac{1}{t} \ . \tag{E.4}$$

In the limit $\tau \to 0$, the exact value of $\Pi *(x,t)$ tends to the conventional value of $\Pi_{ort}(x,t)$, as it should be. However, if $\tau \neq 0$, then for sufficiently large $t$ such that

$$\frac{m\delta x^2}{\hbar\sqrt{t\tau}} << 1 \ , \tag{E.5}$$

we get by virtue of (E.3) and (E.4):

$$\frac{\Pi *(x,t)}{\Pi_{ort}(x,t)} \approx \frac{m\delta x^2}{\tau\hbar} \ . \tag{E.6}$$

Thus, the larger the mass of the particle and/or the broader the initial wave packet, the smaller the ratio of the rates of the wave packet's spreading in the random Gaussian and in the deterministic times. In other words, a massive particle has a "more classical" behavior in the Gaussian time than in the deterministic time.